\documentclass{aa}
\usepackage{epsfig}

\newcommand{\ang}{$\rm \AA$}

\newcommand{\msun}{M$_{\odot}$}

\newcommand{\degree}{$^{\rm o}$}

\newcommand{\msunyr}{\mbox{$M_{\odot} {\rm yr}^{-1}$}}
\newcommand{\kms}{km~${\rm s}^{-1}$}
\newcommand{\ha}{H$\alpha$}

\newcommand{\bb}{\bibitem[]{bla}}

\def\lesssim{\mathrel{\hbox{\rlap{\hbox{\lower4pt\hbox{$\sim$}}}\hbox{$<$}}}}
\def\gtrsim{\mathrel{\hbox{\rlap{\hbox{\lower4pt\hbox{$\sim$}}}\hbox{$>$}}}}

\def\arcsec{\hbox{$^{\prime\prime}$}}

\begin{document}


\title{Resolved polarization changes across \ha\ in the 
classical T Tauri star RY Tau}

\author{Jorick S. Vink{$^1$}, Janet E. Drew{$^1$}, Tim J. Harries{$^2$}, Ren\'e D. Oudmaijer{$^3$}, 
\and Yvonne C. Unruh{$^1$}}
\offprints{Jorick S. Vink, j.vink@ic.ac.uk}
\institute{Imperial College of Science, Technology \& Medicine, Blackett Laboratory, 
           Prince Consort Road, London, SW7 2BZ, U.K.
           \and
           School of Physics, University of Exeter,
           Stocker Road, Exeter EX4 4QL, UK
           \and
           The Department of Physics and Astronomy, 
           E C Stoner Building, Leeds, LS2 9JT, U.K.}

\titlerunning{\ha\ spectropolarimetry of RY Tau}
\authorrunning{Jorick S. Vink et al.}

\abstract{We present linear \ha\ spectropolarimetry of the classical 
T Tauri star RY~Tau. A change in the polarization percentage and position angle across 
\ha\ is detected, which suggests that line photons are scattered in a 
rotating disc. We derive the position angle from the slope 
of the loop in the $(Q,U)$ diagram and find it to be 146 $\pm$ 3\degree. 
This is perpendicular to the position angle of the disc of 48 $\pm$ 5\degree\ 
as deduced from submillimeter imaging by Koerner \& Sargent (1995). 
This finding is consistent, as scattering off the imaged millimeter disc 
is expected to yield a polarization signature in a direction that is 
rotated by 90\degree\ from this disc. 
The observed spectropolarimetric behaviour of RY~Tau is reminiscent of that seen 
in a large group of Herbig Ae stars, suggesting a common circumstellar origin of 
the polarized photons. 
}

\maketitle


\section{Introduction}
\label{s_intro}

Low-mass stars form through the collapse of a rotating interstellar cloud, thereby 
creating a circumstellar disk. During the subsequent  
pre-main sequence (PMS) T Tauri phase circumstellar material accretes 
through the disc onto the central star, most likely via magnetospheric 
funnels (see Johns-Krull et al. 1999 and references therein).
While the basic picture of low-mass star formation is widely accepted, many 
issues remain. For instance, little 
is known about the size of the inner hole, the shape of the inflow and outflow, 
or in short, the geometry of material close to the star. For more massive stars 
the picture becomes even more unclear.
For stars above 10 $M_\odot$ there is not even any consensus on the underlying 
star formation process itself (e.g. Bonnell et al. 1998 versus Behrend \& Maeder 2001).
It is clear that observations of the near-star environment of 
a range of PMS stars are needed to understand star formation as a function 
of mass.

Spectropolarimetry is a powerful tool for studying the near-star regions 
of PMS stars and determining their geometries. The technique has widely been applied 
to early-type stars. Here circumstellar free electrons -- e.g. in a 
disc --  are able to polarize the continuum light more than the line photons, as emission 
lines are formed over a much larger volume (e.g. in a disc) 
than the continuum. Hence, the line 
photons are subject to fewer scatterings leading to a drop in the polarization percentage
across the emission profile (in the absence of interstellar polarization). 
This `line effect' is often referred to as `depolarization' (e.g. Clarke \& McLean 1974). 
The high frequency of depolarization line effects in Herbig Be 
stars (Oudmaijer \& Drew 1999, Vink et al. 2002) provides evidence that 
these stars are surrounded by flattened structures. 
A switch in phenomenology is likely to occur at some point 
working down the stellar mass range, since different physical mechanisms may 
play a role at different spectral types. For instance, radiation 
pressure forces are likely to play a role for the higher-luminosity Herbig Be stars, 
whereas magnetic fields may become more dynamically prominent at A and later type.  
If magnetic fields are present, the inner accretion disc may 
be truncated, and depolarizations would then be absent 
because the inner hole will lead to reduced intrinsic continuum 
polarization. One might therefore anticipate that line effects would be scarce at later 
spectral types.  
However, this is not the case, and the frequency from early B to late A spectral type 
does not drop at all (Vink et al 2002). 
The data on Herbig Ae stars suggest that here the {\it line} itself is polarized 
(rather than the continuum), and that perhaps hot spots on the stellar surface 
provide the required compact source of H$\alpha$ photons that can be  
scattered off material within a rotating disc.

For the even later type (G-K) T~Tauri stars only limited narrow-band
linear polarimetry has been attempted in the past (Bastien \& Landstreet 
1979, Bastien 1982).  In the compilation of Bastien (1982), the results
of observations through a 5~\AA\ wide filter, flanked by contemporary 
broadband 5895~\AA\ and 7543~\AA\ observations, are listed for just 6 T Tauri 
stars.  
Of these objects, RY~Tau is the only example for which a change in 
the line with respect to the broadband continuum was detected.
Partly based on what Bastien (1982) noted as a general absence of polarization changes 
across \ha, he advanced what is now the commonly
accepted view of the origin of T Tauri polarization: namely that 
it is due to scattering off extended dusty envelopes.  

We shall show that the wavelength averaging using even these narrow-band 
H$\alpha$ filters can easily fail to pick up polarization changes that occur on a 
finer wavelength scale of an {\AA}ngstrom or less.  
We have already shown that just such complex changes are commonplace among 
Herbig~Ae stars (Vink et al 2002).  Here we report the 
first spectropolarimetry observation of an object classified 
as a T~Tauri star, and we find similar complexity.
Again this is RY~Tau. We show that if our data were averaged -- as 
in the \ha\ filter observations of Bastien -- they would most likely 
have produced a null result!


In Sects.~\ref{s_obs} and ~\ref{s_results} we discuss how the observations were 
obtained and present the linear \ha\ spectropolarimetry of RY~Tau.
We detect polarization changes and a clear rotation in the polarization 
angle across \ha, which translates into a loop in $(Q,U)$ space, a behaviour 
often seen in Herbig Ae stars. 
In Sect.~\ref{s_disc}, we review other relevant information on RY~Tau (Sect.~\ref{s_ry}), and 
discuss how our data add to our understanding of the complicated flows around this and 
other T Tauri stars (Sect.~\ref{s_specry}). Finally, in Sect.~\ref{s_final}, we 
summarize, and discuss what the RY Tau data may teach about more massive PMS stars in turn.


\section{Observations}
\label{s_obs}

The linear spectropolarimetry for RY~Tau was obtained on the 
night of December 26 (2001) using the ISIS spectrograph 
mounted on the Cassegrain focus of the 4.2-meter William 
Herschel Telescope (WHT), La Palma. A slit width
of $1.1 \arcsec$ was used. 
Although the sky was relatively clear, the seeing was rather poor 
($\sim 2 \arcsec$).
We used a 1024 $\times$ 1024 pixel TEK-CCD detector
with the 1200R grating, which 
yielded spectral coverage from 6370 -- 6760 \ang.
The spectral resolution was approximately 35 \kms\ 
around \ha\ (measured from arc lines).

To obtain the linearly polarized component in the starlight, 
ISIS was equipped with the appropriate polarization optics.
This consisted of a rotating half-wave plate and a calcite block to rotate 
and separate the light into the ordinary (O) and extraordinary (E) light waves.  
Two holes in the Dekker allowed for simultaneous observations 
of the object and the sky.  Hence, in each observation, four spectra were 
recorded: the O and E rays of both the target and sky.  
One complete observation set consisted of a series of four 
exposures at half-wave plate position angles of 
0\degree, 45\degree, 22.5\degree, and 67.5\degree.
After each of these sets of four frames was obtained, the target 
and the sky were interchanged by placing the star in the other Dekker 
hole, so as to compensate for any possible asymmetry in the detector 
or the instrument. The total exposure time on RY~Tau was 2640 seconds.   
Polarized and zero-polarization standards were also observed during the run 
and an intrinsic instrumental polarization of 0.01 percent and a PA of 
31\degree\ (the PA is extremely uncertain) were revealed. As the instrumental 
polarization is found to be negligible, we have not corrected for it.


\begin{figure}
\centerline{\psfig{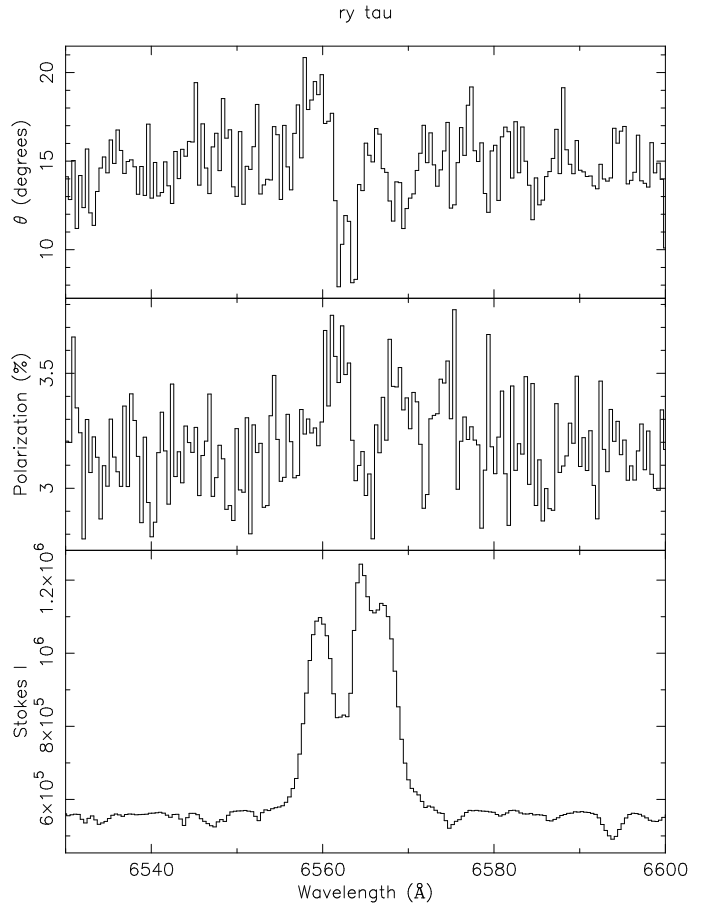}}
\caption{Triplot of the observed polarization spectra of the classical T~Tauri star RY~Tau.
The Stokes $I$ `intensity' spectrum is shown in the lowest panel of the triplot, 
the \%Pol is indicated in the middle panel, while the PA ($\theta$; see Eq.~2) is 
plotted in the upper panel. The data are not rebinned.}
\label{f_rytau}
\end{figure}

The data reduction steps included bias-subtraction and flat-fielding, cosmic ray 
removal, extraction of the spectra, and wavelength calibrations of the O and 
E spectra. The Stokes parameters $Q$ and $U$ were subsequently 
determined from these O and E rays, leading to the percentage linear 
polarization $P$ and its Position Angle (PA) or $\theta$:

\begin{equation}
P~=~\sqrt{(Q^2 + U^2)}
\end{equation}
\begin{equation}
\theta~=~\frac{1}{2}~\arctan(\frac{U}{Q})
\end{equation}
The achieved (relative) accuracy of the polarization data is in principle only 
limited by photon-statistics and can be very small (typically 0.01 \%). 
However, the quality and the amount 
of data taken on spectropolarimetric standard stars is at present not 
yet sufficient to guarantee absolute accuracies to within 0.1\%. 


\section{Results}
\label{s_results}

\begin{figure}
\centerline{\psfig{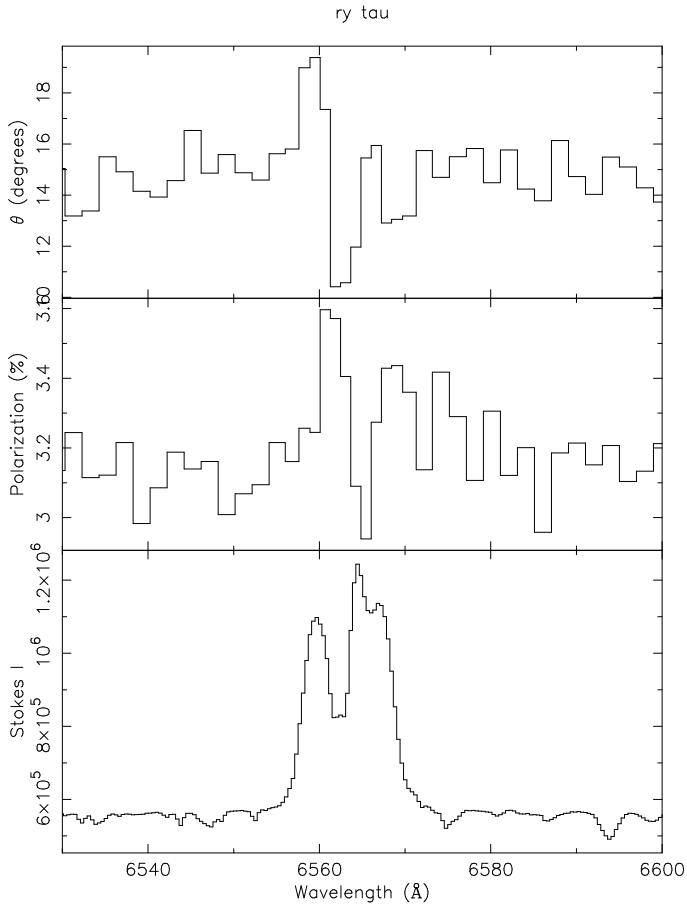}}
\caption{Same as Fig.~\ref{f_rytau} but now rebinned such 
that the 1$\sigma$ error in the polarization corresponds to 0.09\% 
as calculated from photon statistics.}
\label{f_rytaubin}
\end{figure}

\begin{figure}
\centerline{\psfig{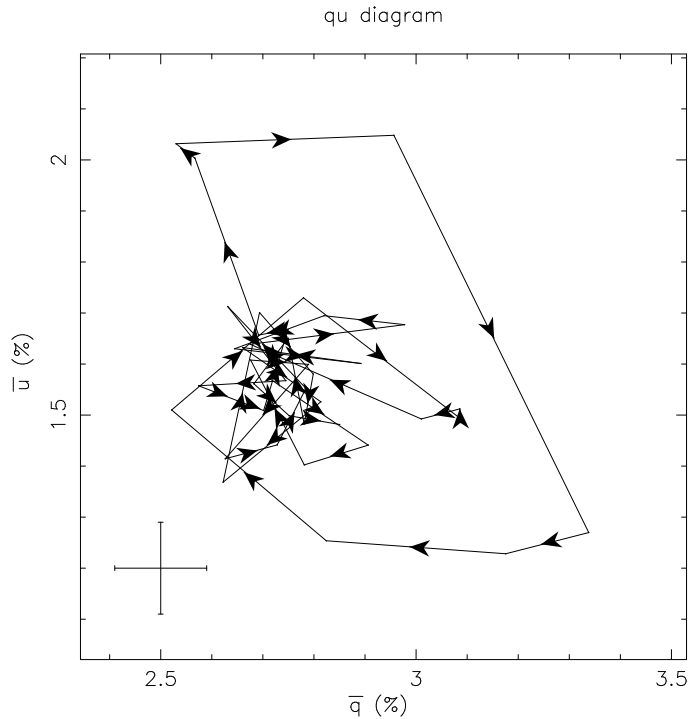}}
\caption{$(Q,U)$ diagram of RY~Tau, also rebinned to a 1$\sigma$ error in the 
polarization of 0.09\%. Note that the direction of the arrow denotes the increase 
in wavelength.} 
\label{f_rytauqubin}
\end{figure}

The linear spectropolarimetry data for RY~Tau are shown in 
the triplot of Fig.~\ref{f_rytau}. The Stokes $I$ `intensity' spectrum is depicted  
in the lowest panel, the \%Pol is indicated in the middle panel, while the PA ($\theta$; see 
Eq.~2) is plotted in the upper panel. 
Note the presence of clear changes across the \ha\ line in both the \%Pol and the PA. 
The Stokes $I$ \ha\ profile itself is complex, as is common for T Tauri stars. One interpretation
is that it consists of both a double-peaked profile, which is consistent (but no proof) 
of rotation in a disc, as well as a P~Cygni component on top of this, indicative of a wind.
As far as the spectropolarimetry is concerned, we choose to rebin the data, aiming 
to obtain a reduced and fixed error per bin. However, we need to ensure 
simultaneously that we do not lose relevant spectral resolution as a result of this. 
We found that a constant error of 0.09\% polarization represents the 
best compromise between minimizing the error per bin and resolving the line profile. The 
resulting \ha\ polarization data are shown in the triplot of Fig.~\ref{f_rytaubin}.
The data are alternatively represented in a $(Q,U)$ diagram (Fig.~\ref{f_rytauqubin}) where 
the direction of the arrow denotes the increase in wavelength.
The dense knot of points located at $(Q_{\rm obs},U_{\rm obs})=(2.75, 1.56)\%$, represents 
the observed level of continuum polarization with a mean value $P = 3.16 \pm 0.01 \%$ and 
a PA of 14.2 $\pm$ 0.1\degree. Note the clear presence of the loop, which is the equivalent of the 
flip in the PA across \ha\ in the upper panel of Fig.~\ref{f_rytaubin}. This would not 
have been observed when a 5 \AA\ narrow-band filter (as used by Bastien) had 
been used. Integrating the polarization over 5 \AA\ from 6560 -- 6565 \AA\ yields a 
mean PA of 14\degree, which would be indistinguishable from the continuum PA of 14.2\degree.
Yet we {\it do} detect structure of less than an {\AA}ngstrom across \ha\ by 
performing spectropolarimetry.

If the scattering leading to polarization occurs predominantly in a 
rotating disc-like configuration, the subsequent breaking of left-right 
reflection symmetry in the velocity field leads to a changing PA 
with wavelength (see e.g. McLean 1979, Wood et al. 1993), which 
appears as a `loop' in $(Q,U)$ space. This is precisely what is 
detected here in the classical T~Tauri star RY~Tau and has previously been 
seen in a large group of Herbig Ae stars (Vink et al. 2002). 


\section{Discussion}
\label{s_disc}

\subsection{RY~Tau}
\label{s_ry}

RY~Tau is a classical T Tauri star of spectral type F8III (Mora et al. 
2001). Photometric monitoring studies have shown that the brightness of the star
varies by up to 2-3 mag over a timescale of decades (e.g. Beck \& Simon 2001 and 
references therein). Herbst et al. (1994) classify the object as a type III variable. 
Although there is no consensus on the underlying mechanism of these objects, variable circumstellar 
obscuration most likely plays a role, as in UX~Ori variables.  
The star has a rather low level of veiling, i.e. $\le$ 0.1 
in the visible (e.g. Basri et al. 1991), and a rotational velocity of $v \sin i = 
55 \pm 3$ \kms\ (Mora et al. 2001). The mass-accretion rate is of the order of 
$\dot{M}$ = 7.5 $\times$ $10^{-8}$ $\msunyr$ (Bertout et al. 1988). Spectroscopic 
variability has also been noted, but whether the \ha\ line strength correlates with the
$V$ magnitude (Vrba et al. 1993), or not (Petrov et al. 1999) is under debate.

Most relevant for this study is the high level of 
linear polarization of a few percent. The variability of its polarization
was discovered by Vardanyan (1964) and has been confirmed ever since (e.g. Serkowski 1969, 
Schulte-Ladbeck 1983, Bastien 1985, Oudmaijer et al. 2001). 
The linear percentage polarization \%Pol has been reported to vary 
between 1 and 6 \%, and its PA between $-$20 and 75\degree, with \%Pol 
$\simeq 3$ and PA $\simeq$ 20\degree\ being typical `median' 
values (taken over several epochs and with values reported by authors employing  
different techniques).  

The interpretation of the level of polarization as well as its 
variability is ambiguous for this, as for any other, T Tauri star.
The scattering that leads to polarization could either be due to gas or dust, and
although the wavelength dependence of the observed polarization surely 
implies a strong dust component, this dust polarization  
could be interstellar, as well as circumnebular, and/or circumstellar:
decomposing these different contributions has since long been recognized 
to be a particularly difficult task (see e.g. McLean \& Clarke 1979). 
The variability of the observed polarization may help identify its 
origin. It has been suggested that rotational modulation 
may signal cool and hot spots on the stellar surface (e.g. Stassun \& Wood 1999), but 
also variable extinction by a dusty disc can lead to quasi-periodic behaviour 
of the observed level of dust polarization (e.g. M\'enard \& Bastien 1992).

\subsection{Spectropolarimetry on RY~Tau}
\label{s_specry}

Spectropolarimetry is unique in that it is possible to retrieve the PA from $(Q,U)$ plane 
excursions without relying on uncertain values of the foreground polarization. This is because
interstellar dust treats the line and the continuum in an entirely similar way. 
For instance, Oudmaijer et al. (1998) were able to obtain 
the intrinsic PA of the scattering material in the rotating disc of the B[e] star HD~87643.
It has been suggested that the intrinsic PA of the scattering material can 
be obtained from the increased polarization across blueshifted absorption (McLean 1979). 
If one accepts the view that the Stokes $I$ profile in Figs.~\ref{f_rytau} and \ref{f_rytaubin} 
indeed consists of a double-peaked as well as a P~Cyg profile, one may 
obtain the intrinsic PA from the increased polarization over the absorption part of the line, 
which occurs at a PA of 160 -- 167\degree.
Alternatively, we can measure the PA of the scattering material 
from the slope of the loop in the $(Q,U)$ diagram, $\theta_{\rm intr}$ = 0.5 
$\times$ atan($\Delta U/\Delta Q) \simeq$ 146 $\pm$ 3\degree. Note that the error on 
this PA is determined as follows. We have replotted the polarization 
data in $(Q,U)$ space applying a range of sampling errors. At each replotting, the slope of 
the $(Q,U)$ loop was measured. The error quoted above is the standard deviation of these 
10 independent measurements.

Is there any other information available than just the PA of the $(Q,U)$ excursions?
Yes. From the observed level of continuum polarization in these new data  with  
$(Q_{\rm obs},U_{\rm obs})=(2.75, 1.56)\%$ (see Sect. \ref{s_results}), we may be able 
to retrieve the level of intrinsic continuum polarization from the 
vector addition in $(Q,U)$ space, {\it if} we are able to find a reliable value for the ISP.
Both Efimov (1980) and Petrov et al. (1999) derive the ISP toward RY~Tau obtaining values of 
$P=2.69\%$, $\theta =$ 27\degree, and $P=2.84\%$, $\theta =$ 26\degree\ respectively. Although 
Efimov used field stars, while Petrov et al. used the variability of a huge body of 140 
datapoints of the star itself, these values are strikingly alike, and may give some confidence
as being close to the true value. 
Adopting the Petrov et al. (1999) value for the ISP (at 5500 \AA), or alternatively 
$(Q_{\rm ISP},U_{\rm ISP})$ = (1.75, 2.24)\%, we can estimate the intrinsic continuum 
polarization degree of RY~Tau. Employing the vector relationship 
$(Q_{\rm intr},U_{\rm intr}) = (Q_{\rm obs},U_{\rm obs}) - (Q_{\rm ISP},U_{\rm ISP})$, we 
find: (2.75, 1.56) - (1.75, 2.24) = (1.00, $-$0.68), This corresponds 
to an {\it intrinsic} level of polarization of 1.21\% at a PA of 163\degree. 
The statistical error in this PA is insignificantly small, but the error in subtracting 
the ISP is subject to systematic error. Nonetheless, assuming that the estimate 
for the ISP toward RY~Tau is correct, the PA of 163\degree\ derived this way 
agrees very well with the PA from the blueshifted absorption, and is only 18\degree\ off 
from the intrinsic $\theta_{\rm intr}$ of 146 $\pm$ 3\degree\ as found from the the slope of 
the $(Q,U)$ loop. 

How does all this compare to other evidence for discs around RY~Tau?
Koerner \& Sargent (1995) detected gaseous emission around RY~Tau at millimeter (CO 2-1) 
wavelengths along a direction of elongation with a PA of $\simeq 48 \pm 5$\degree\  
(where the formal error is based on the goodness of fit only). In addition
velocity gradients parallel to this elongation suggest that the material is rotating
in a Keplerian disc. 
If our detected change in PA across \ha\ is due to scattering in a rotating 
circumstellar disc, as the loop in $(Q,U)$ suggests, one would expect 
the PA derived from the polarimetry to be perpendicular to the CO disc, 
i.e. at a PA of $\simeq 48 + 90$ = 138\degree\ (assuming multiple scattering 
effects are not important). To within the errors this agrees with 
the slope of the loop we found in $(Q,U)$ space. 
It is important to note that the deconvolved size at mm wavelengths for RY~Tau is less 
than the full-width at half-maximum beam size, and hence that the error on 
$\theta_{\rm CO}$ is likely to be greater than the quoted value of 5\degree\ (Koerner \& 
Sargent 1995). 

\section{Final Remarks}
\label{s_final}

To summarize, both the \ha\ spectropolarimetry with a PA of $\simeq$ 146\degree, as 
well as the ISP corrected continuum measured PA of $\simeq$ 163\degree, are 
considered to be consistent with the mm detection of the 
rotating disc around RY~Tau with a PA of $\simeq$ 48\degree. Because the  
scattering off a disc is expected to yield a polarization signature in a 
direction that is rotated by 90\degree\ from the disc, a rather 
consistent picture of the RY~Tau disc system emerges. 

Traditionally, the switch between early and late type stars in the Hertzsprung-Russell Diagram is 
believed to occur between spectral type A and F, as this is where the cooler stars have convective, while 
hotter stars possess radiative outer envelopes. Similarly, the switch between low and high-mass star 
formation may be thought to occur at the T~Tauri/Herbig Ae boundary 
(at $\simeq$~2 \msun), where magnetic fields may be 
of dynamical importance in the T Tauri stars, but not in the Herbig Ae objects. 
The striking resemblance in spectropolarimetric behaviour 
between the T Tauri star RY~Tau and a large group of Herbig Ae stars, as signaled by the detected 
change in PA and the $(Q,U)$ loop, suggest a common origin of the polarized line photons, and hint that 
low and high mass pre-main sequence stars may have more in common than had hitherto been suspected.


\begin{acknowledgements}
We thank the referee, Pierre Bastien, for his constructive comments.
The allocation of time on the William Herschel Telescope was awarded 
by PATT, the United Kingdom allocation panel.  JSV is funded by the 
Particle Physics and Astronomy Research Council of the United Kingdom.  
The data analysis facilities are provided by the Starlink Project, which 
is run by CCLRC on behalf of PPARC. 
\end{acknowledgements}

\end{document}